# Mathematical Modeling of Aerodynamic "Space -to - Surface" Flight with Trajectory for Avoid Intercepting Process

Serge Gornev

Aerospace CyberResearch Co.

E-mail: sgornev@hotmail.com

## Abstract

Modeling has been created for a "Space-to-Surface" system defined for an optimal trajectory for targeting in terminal phase with avoids an intercepting process. The modeling includes models for simulation atmosphere, speed of sound, aerodynamic flight and navigation by an infrared system. The modeling and simulation includes statistical analysis of the modeling results.

Aerospace CyberResearch Company has created modeling and simulation of aerodynamic flight. The flight has an optimal trajectory for targeting and may be used for civil and military missions. [1]

The scenario, which uses the head of the "Space-to-Surface" system with optimal trajectory, includes:

· Launching the head from a space orbit with an altitude of (H=250-300km) by pulse from a space platform or maybe launching it by ballistic missile. In addition, the head may be launched from an aircraft in the "Space-to-Surface" mission during space flight.

· After making the pulse from the space platform the head enters the atmosphere. In next phase, it maneuvers in the atmosphere taking on horizontal flight. While in horizontal flight the head is searching for the target by the infrared guidance system. Automatic control of the flight has the following conditions of altitude:

¨ Maneuver in atmosphere made with radius R=35-40km.
¨ Horizontal flight must have altitude H= 33-40 km for searching target.

The time for space flight is T=15-20 min until the time the head enters the atmosphere. The time T=15-20 min is needed for the flight from space to enter to atmosphere with angle of attack A=3-4°.

Trajectory of the head in the atmosphere includes four phases:

1. Aerodynamic flight with the included entrance to the atmosphere with an angle of attack A=3-4°. This part of the trajectory has an altitude of H=90-100 km and a velocity of V=7.6 km/sec.

2. Maneuver in atmosphere from H=80-90 km to H=30-40 km in altitude. The velocity in this phase is V=5-7.6 km/sec.

3. Target searching while in horizontal flight. The velocity in this phase is V=3-5 km/sec.

4. The terminal phase, navigation and deployment of the target. The velocity in this phase is V=2.8-3 km/sec.

All phases of the flight simulation include automatic control of the flight using parameter U. The mathematical model includes differential equations with parameter U integrated in four phases of flight. In the first phase it has been used for was gravitational flight:

1. $U = \cos\theta$, where $\theta$- angle between horizontals and vector of head velocity.

In second phase it has been used for the maneuver in atmosphere with radius R=45 km:

2. $U = V^2/gR + \cos\theta$, where V- velocity of head.

In third phase it has been used for when the flight takes on horizontal flight with searching target:

3. $U = k(H-Y) + \cos\theta$, where k – proportional coefficient, H=35 km, Y –current altitude.

In fourth terminal phase it has been used for the infrared guidance to the target:

4. $U = k\varphi$, where $\varphi$ - angle between head and target.

In terminal phase the design of head may include the engine and make accelerating of velocity on final phase of flight. The head in terminal phase may make maneuvers for avoid intercepting processing.

The time for flight in the atmosphere with an altitude of H=100 km to deploy the target is T=60-90 sec with a horizontal range of X=500-625 km.

The head, searching for the target, starts in phase 3 by the infrared guidance system or may using the GPS for targeting. The head may use the design and the technology of cruise missile.

This simulation has been created by the integration of differential equations and includes modeling of the head flight; searching, guidance, and automatic control and deploy the target.

As stated the modeling includes a models of atmosphere, speed of sound, for automatic control of flight and for an infrared guidance. The simulation has been created in the programming language Fortran. The modeling includes statistical analysis of the modeling results. The simulation produced results with high probability of targeting inaccuracy (with an range error of R=3-10m).

The simulation of the head is used with a few parameters:

- Weight m=1450-1550 kg
- Aerodynamic coefficient K=CY/CX~2
- Wing area S~2m²

Gravitational flight, maneuver in the atmosphere, horizontal flight, searching, guidance and deployment the target are defined in the models differential equations. All parameters (X, Y, Z, V, T, U, A) of the flight are integrated, where:

- X, Y, Z- coordinates of head;
- T-time of flight;
- V- velocity of head;
- U-parameter of automatic control;
- A-angle of attack.

**Conclusion**

The simulation was calculated with two-dimensional and three-dimensional flight, design and analysis. The modeling and simulation was created using various software applications describing all parameters. The trajectory which is includes four phases of flight may avoid any intercepting process. The guidance created for each part of flight and includes targeting in very short time in atmosphere, less than T < 90sec. In conclusion, the modeling has statistic definitions and systems analysis of the modeling results with recommendations for design of a head.